\documentclass{article}%
\usepackage[top=3cm, bottom=3cm, left=3cm, right=3cm]{geometry}
\usepackage{amsmath}
\usepackage{amsfonts}
\usepackage{amssymb}
\usepackage{graphicx}%
\begin{document}

\title{$f_{2}(1270),$ $a_{1}(1260)$ and $f_{0}(1370)$ as dynamically reconstructed
quark-antiquark states\thanks{Based on the contribution given at the Chiral10
Workshop, Valencia (Spain), June 21-24, 2010.}}
\author{Francesco Giacosa\\\emph{Institut f\"{u}r Theoretische Physik Universit\"{a}t Frankfurt}\\\emph{\ Johann Wolfgang Goethe - Universit\"{a}t Max von Laue--Str. 1}\\\emph{\ D-60438 Frankfurt, Germany}}
\maketitle

\begin{abstract}
It is explained why the interpretation of the resonances $f_{2}(1270),$
$a_{1}(1260)$ and $f_{0}(1370)$ as quark-antiquark states is legitimate. The
result of the quark model and of recently performed Bethe-Salpeter studies are
not (necessarily) in conflict and can be understood as two different
approaches toward the description of the same quark-antiquark
resonances.\newline\newline

\textbf{Pacs:} 12.39.Mk,11.10.St,11.15.Pg,11.30.Rd


\end{abstract}

\section{Introduction}

The interpretation of hadronic resonances is an important subject of
low-energy hadron physics. Can all resonances be described in terms of the
building blocks of QCD, quarks and gluons? Are there, on the contrary,
`dynamically generated' resonances, which emerge as molecular bound states
upon interactions of other, more fundamental hadrons?

In this work we shall concentrate on some particular mesons in the mass range
1-1.5 GeV: $f_{2}(1270),$ $a_{1}(1260)$ and $f_{0}(1370).$ These states were
investigated both in the old-fashioned quark model \cite{quarkmodel} and in
more recent studies \cite{refs}. In the quark model they are interpreted as
quark-antiquark pairs: $f_{2}(1270)\equiv\bar{n}n\equiv\sqrt{1/2}(\bar
{u}u+\bar{d}d)$ with quantum numbers $J^{PC}=2^{++},$ $a_{1}^{0}(1260)\equiv$
$\sqrt{1/2}(\bar{u}u-\bar{d}d)$ with quantum numbers $J^{PC}=1^{++}$
(similarly for the other charged states), $f_{0}(1370)\equiv\bar{n}%
n\equiv\sqrt{1/2}(\bar{u}u+\bar{d}d)$ with quantum numbers $J^{PC}=0^{++}.$
Indeed, the quarkonium assignment describes very well the masses and also the
decays properties for the whole tensor meson nonet $J^{PC}=2^{++}$
\cite{tensormio}, to which $f_{2}(1270)$ belongs. It also functions well for
the axial-vector meson nonet $J^{PC}=1^{++},$ to which $a_{1}(1260)$ belongs.
The scalar mesons, such as $f_{0}(1370),$ are as usual controversial objects
in QCD, however the interpretation of $f_{0}(1370)$ as predominantly
quarkonium is in agreement with the present results, see Refs.
\cite{scalars,denis} and refs. therein for the presentation of various mixing patterns.

In the works of Refs. \cite{refs} the very same resonances $f_{2}(1270),$
$a_{1}(1260)$ and $f_{0}(1370)$ have been obtained in -at first sight- utterly
different light: the resonances $f_{2}(1270)$ and $f_{0}(1370)$ are
interpreted as $\rho\rho$ molecular states, and $a_{1}(1260)$ is interpreted
as $\rho\pi$ molecular state. These works are based on the Bethe-Salpeter (BS)
equation applied to mesonic low-energy chiral Lagrangians describing $\rho
\rho$ and $\rho\pi$ interactions.

The two descriptions seem mutually exclusive: a mesonic molecular state is a
different object than a quark-antiquark bound state. This is surely true, but
a closer look at the problem is necessary. Namely, to which extent can one
conclude that these resonances are hadronic molecular states? Surely, the BS
equation represents a well-defined field theoretical framework to describe
bound states. However, the BS approach is used in the context of low-energy
hadronic theories as a unitarization method: the BS-resummation scheme is
applied to a low-energy effective Lagrangians with a limited range of
validity. The masses of the poles of BS amplitudes lie above the validity of
the corresponding low-energy hadronic theory. This is indeed a subtle point
that requires a careful study, which we will present in this paper along the
line of Ref. \cite{dynrec}.

It is useful to discuss the problem with the help of two examples:

(a) Positronium states in QED: the QED Lagrangian contains two fields, the
photon and the electron. Further composite states with a mass of about
$2m_{e}$ emerge upon electron-positron interactions. These states are the
positronia, i.e. bound states of electron and positron due to photon exchange.
Positronia are genuine dynamically generated states of molecular type which do
not appear in the original QED Lagrangian.

(b) Fermi-Lagrangian and the nature of the $W$ meson: in the standard model
(SM) the fields $W,$ electron and neutrino are elementary. The mass of the
weak $W$ boson is however very large (80 GeV). When integrating out from the
SM the $W$-field, the Fermi Lagrangian for $e$-$\nu$ interaction emerges. If
one applies unitarization techniques by resumming $e$-$\nu$ loops to the Fermi
Lagrangian, the existence of the $W$ meson can be inferred. However, this does
not mean that the $W$ meson is a `dynamically generated' bound state of an
electron and a neutrino. Indeed, the $W$ meson is exactly as elementary as $e$
and $\nu.$ One can rather say that the $W$ meson can be `reconstructed' by
unitarizing the low-energy Fermi Lagrangian.

The question concerning the resonances $f_{2}(1270),$ $a_{1}(1260),$ and
$f_{0}(1370)$ can be summarized as follows: are they analogous to the case (a)
or the case (b)? In this work we argue that they are analogous to the case
(b). This means that these resonances are not hadronic molecular bound states,
but rather standard quarkonia. They are obtained upon unitarizations of
low-energy hadronic Lagrangians, just as the $W$ meson can be obtained from
the low-energy Fermi Lagrangian.

\section{Dynamical reconstruction of quark-antiquark states}

\subsection{A short survey of low-energy theories of QCD}

The QCD Lagrangian $\mathcal{L}_{QCD}$ contains quarks and gluons, which due
to confinement are not the relevant degrees of freedom at low energy. The
proper degrees of freedom are colorless hadron states. The effective
Lagrangian describing these hadrons up to (some) maximal energy $E_{\max}$ is
denoted as $\mathcal{L}_{had}(E_{\max},N_{c}).\ $

We briefly describe four important particular cases of $\mathcal{L}%
_{had}(E_{\max},N_{c}).$

(i) The case $E_{\max}\simeq2$ GeV is interesting from a phenomenological
point of view, because all the low-lying nonets ((pseudo)scalars and
(axial)vectors) lie below this energy. Unfortunately, $\mathcal{L}%
_{had}(E_{\max}\simeq2GeV,N_{c}=3)$ is unknown. It is in fact not possible to
derive it from $\mathcal{L}_{QCD}.$ (For recent attempts to describe -part of-
$\mathcal{L}_{had}(E_{\max}\simeq2$ GeV$,N_{c}=3)$ including (pseudo)scalar
and (axial)vector mesons see Ref. \cite{denis}).

(ii) When setting $E_{\max}=E_{\chi PT}\simeq300$ MeV, the Lagrangian
$\mathcal{L}_{had}(E_{\max}\simeq300$ MeV$,$ $N_{c}=3)$ contains only the 3
light pions:
\begin{equation}
\mathcal{L}_{\chi PT}=\mathcal{L}_{had}(E_{\max}=E_{\chi PT},N_{c}%
=3)=\sum_{k=1}^{3}\left[  \frac{1}{2}\left(  \partial_{\mu}\pi_{k}\right)
^{2}-\frac{1}{2}M_{\pi}^{2}\pi_{k}^{2}\right]  +\mathcal{L}_{int}^{\pi}\text{
,}%
\end{equation}
whereas $\mathcal{L}_{int}^{\pi}$ describes the interaction term. This is
indeed the Lagrangian of chiral perturbation theory \cite{chpt}, whose terms,
but not the related coupling constants (known as low-energy coupling
constants, LECs), can be determined by considerations based on chiral
symmetry. Note, if $\mathcal{L}_{had}(E_{\max}\simeq2GeV,N_{c}=3)$ were known
it would be possible, upon integrating out all the heavier fields, to
determine exactly $\mathcal{L}_{\chi PT}$: both the operators and the LECs
would be calculable. In the next subsection a toy model where this operation
is possible is shown.

(iii) If, instead, we chose $E_{\max}\simeq1$ GeV we obtain the effective
Lagrangian
\begin{equation}
\mathcal{L}_{had}(E_{\max}\simeq1\text{ GeV,}N_{c}=3)=\mathcal{L}_{\chi
PT+VM}\text{ ,}%
\end{equation}
i.e. a Lagrangian which describes the pseudoscalar mesons, the vector mesons
and their interactions \cite{chptvm}. It is out of this Lagrangian that the
resonances $f_{2}(1270),$ $a_{1}(1260),$ $f_{0}(1370)$ where obtained in Ref.
\cite{refs} upon unitarization based on the BS-equation.

(iv) There is one theoretical limit in which $\mathcal{L}_{had}(E_{\max}%
,N_{c})$ can be determined: the large-$N_{c}$ limit. In fact, for $N_{c}>>1$
the theory $\mathcal{L}_{had}(E_{\max},N_{c}>>1)$ contains only free quarkonia
and glueballs with a mass below the maximal energy $E_{\max}$:%
\begin{equation}
\mathcal{L}_{had}(E_{\max},N_{c}>>1)=\sum_{k=1}^{N_{\overline{q}q}}\left[
\frac{1}{2}\left(  \partial_{\mu}\phi_{k}\right)  ^{2}-\frac{1}{2}%
M_{_{\overline{q}q},k}^{2}\phi_{k}^{2}\right]  +\sum_{h=1}^{N_{gg}}\left[
\frac{1}{2}\left(  \partial_{\mu}G_{h}\right)  ^{2}-\frac{1}{2}M_{G,h}%
^{2}G_{h}^{2}\right]  \text{ .}%
\end{equation}

\subsection{A toy-model and its analogy with the hadronic world}

In order to explain the issue it is useful to introduce a simple toy-model
with the scalar fields $\varphi$ and $S$ \cite{dynrec,lupo}:
\begin{equation}
\mathcal{L}_{\text{toy}}(E_{\text{max}},N_{c})=\frac{1}{2}\left(
\partial_{\mu}\varphi\right)  ^{2}-\frac{1}{2}m^{2}\varphi^{2}+\frac{1}%
{2}\left(  \partial_{\mu}S\right)  ^{2}-\frac{1}{2}M_{0}^{2}S^{2}%
+g(N_{c})S\varphi^{2}-\frac{g(N_{c})^{2}}{2M_{0}^{2}}\varphi^{4}\text{ ,}%
\end{equation}
where the large-$N_{c}$ dependence is expressed via the scaling $g(N_{c}%
)=g_{0}\sqrt{3/N_{c}}$. The two masses are large-$N_{c}$ independent and the
decay width $S\rightarrow2\varphi$
\begin{equation}
\Gamma_{S\rightarrow\varphi\varphi}=\frac{\sqrt{\frac{M_{0}^{2}}{4}-m^{2}}%
}{8\pi M_{0}^{2}}\left[  \sqrt{2}g(N_{c})\right]  ^{2}%
\end{equation}
scales as $1/N_{c},$ exactly as if $\varphi$ and $S$ were quark-antiquark
states. (We assume that $M_{0}>2m$ and that the validity of the Lagrangian
$\mathcal{L}_{\text{toy}}$ is such that $E_{\max}>>M_{0}$).

Let us now turn to the determination of an effective low-energy model of this
simplified system. We integrate out $S$ and obtain a low-energy Lagrangian
$\mathcal{L}_{\text{le}}$ valid in the interval $E_{\text{le}}\lesssim
2m<M_{0}$ and depending $\emph{only}$ on the field $\varphi$:
\begin{equation}
\mathcal{L}_{\text{le}}(E_{\text{le}},N_{c})=\frac{1}{2}\left(  \partial_{\mu
}\varphi\right)  ^{2}-\frac{1}{2}m^{2}\varphi^{2}+V,\text{ }V=\sum
_{k=1}^{\infty}V^{(k)}\text{ ,} \label{lle}%
\end{equation}%
\begin{equation}
\text{ }V^{(k)}=L^{(k)}\varphi^{2}\left(  -\square\right)  ^{k}\varphi
^{2},\text{ }L^{(k)}=\frac{g(N_{c})^{2}}{2M_{0}^{2+2k}}\text{ .} \label{lk}%
\end{equation}

It is easy to establish an analogy of the toy-model with the real hadronic
world, see Table 1. $\mathcal{L}_{\text{toy}}(E_{\text{max}},N_{c})$
corresponds to the (unknown) hadronic Lagrangian $\mathcal{L}_{had}(E_{\max
}\simeq2$ GeV$,N_{c}),$ while the low-energy Lagrangian $\mathcal{L}%
_{\text{le}}(E_{\text{le}},N_{c})$ corresponds to a low-energy hadronic
Lagrangian, such as $\mathcal{L}_{\chi PT}$ or $\mathcal{L}_{\chi PT+VM}.$

\begin{center}
\textbf{Table 1}: Analogy%

\begin{tabular}
[c]{|l|l|}\hline
Toy-Model & Hadronic world\\\hline
$\mathcal{L}_{\text{toy}}(E_{\text{max}},N_{c})$ & $\mathcal{L}_{had}(E_{\max
}\simeq2$ GeV$,N_{c})$\\\hline
$\mathcal{L}_{\text{le}}(E_{\text{le}},N_{c})$ & $\mathcal{L}_{\chi PT}\ $or
$\mathcal{L}_{\chi PT+VM}$\\\hline
\end{tabular}

\end{center}

There is however a crucial difference: while in the toy-model the knowledge of
the `full Lagrangian' $\mathcal{L}_{\text{toy}}(E_{\text{max}},N_{c})$ allows
to determine the low-energy counterpart $\mathcal{L}_{\text{le}}(E_{\text{le}%
},N_{c})$ precisely up to an arbitrary order $n$ (see Eq. (\ref{lle})), the
Lagrangians $\mathcal{L}_{\chi PT}$ and $\mathcal{L}_{\chi PT+VM}$ are only
partially known. The terms are determined via symmetry considerations, but the
corresponding coupling constants (LECs), which are analogous to the $L_{k}$ in
Eq. (\ref{lk}), cannot be calculated: they must be obtained via comparison
with experiments. This fact represents also a practical limit of low-energy
effective theories: although it is in principle possible up to work at each
$n$, the technical difficulty due to the fast increasing number of terms and
the large number of unknown related LECs render the calculations doable only
up to the third order.

\subsection{The concept of dynamical reconstruction}

In the framework of the toy-model, the $T$-matrix for $\varphi\varphi$
scattering in the $s$-channel can be calculated from the Lagrangian
$\mathcal{L}_{\text{toy}}(E_{\text{max}},N_{c})$ (at 1-loop, see the first and
the second rows of Fig.1):%
\begin{equation}
T(p^{2})=\frac{1}{-K^{-1}+\Sigma_{\Lambda}(p^{2})},\text{ }K=\frac{(\sqrt
{2}g)^{2}}{M_{0}^{2}-p^{2}}-\frac{(\sqrt{2}g)^{2}}{M_{0}^{2}}\text{ ,}
\label{t-exakt}%
\end{equation}
where $\Sigma_{\Lambda}(p^{2})$ is the loop function, which depends on a
cutoff $\Lambda,$ see Ref. \cite{lupo} for details. This is for our purposes
the `exact' $T$-matrix of the problem \footnote{Clearly this form of the
$T$-matrix is valid in the 1-loop approximation. Even this simple QFT is not
exactly solvable. Neverhteless, the resummed 1-loop expression is regarded as
`exact' in comparison to the approximated BS-form derived later.}.%

\begin{figure}
[ptb]
\begin{center}
\includegraphics[
height=3.4126in,
width=4.0283in
]%
{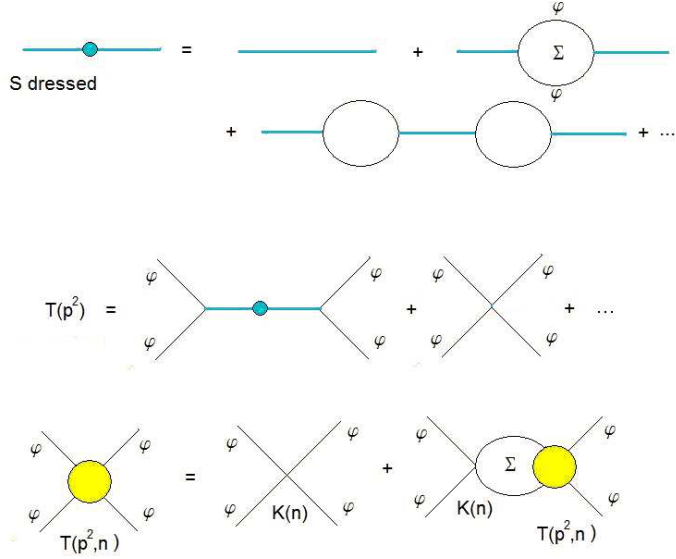}%
\caption{First row: Dressing of $S$-field through loops of $\varphi$-fields.
Second row: Pictorial representation of the 1-loop resummed $T$-matrix
$T(p^{2})$ of Eq. (\ref{t-exakt}). Third row: Pictorial representation of the
BS-approximation $T_{BS}(p^{2},n)$ of Eq. (\ref{tbs}).}%
\end{center}
\end{figure}

Let us now consider the low-energy Lagrangian $\mathcal{L}_{\text{le}%
}(E_{\text{le}},N_{c}),$ in which the potential\emph{\ }$V$ is approximated at
the order $n$: $V(n)=\sum_{k=1}^{n}V^{(k)}$. We can apply a BS-study to this
system, see the third row of Fig.1, obtaining upon resummation the following
approximated form of the $T$-matrix:%
\begin{equation}
T_{BS}(p^{2},n)=\frac{1}{-K(n)^{-1}+\Sigma_{\Lambda}(p^{2})},\text{
}K(n)=\frac{(\sqrt{2}g)^{2}}{M_{0}^{2}}\sum_{k=1}^{n}\left(  \frac{p^{2}%
}{M_{0}^{2}}\right)  ^{k}. \label{tbs}%
\end{equation}
The quantity $K(n)$ is the perturbative amplitude calculated from
$\mathcal{L}_{\text{le}}(E_{\text{le}},N_{c})$ as sum of the first $n$ terms.
Clearly, $T_{BS}(p^{2},n)$ is an approximated function of $T(p^{2})$ of Eq.
(\ref{t-exakt}). The larger $n$, the better is the approximation. Formally:
$\lim_{n\rightarrow\infty}$ $T_{BS}(p^{2},n)=T(p^{2})$.

Now, let us suppose that the low-energy Lagrangian $\mathcal{L}_{\text{le}%
}(E_{\text{le}},N_{c})$ is known, while the original Lagrangian $\mathcal{L}%
_{\text{toy}}(E_{\text{max}},N_{c})$ is unknown. (This is indeed the case of
the real hadronic world, where only $\mathcal{L}_{\chi PT}\ $or $\mathcal{L}%
_{\chi PT+VM}$ are known, but not $\mathcal{L}_{had}(E_{\max}\simeq2$
GeV$,N_{c})$). Moreover, we concentrate on the usually considered case in the
literature: only the first term in the expansion of $\mathcal{L}_{\text{le}%
}(E_{\text{le}},N_{c})$ is kept. This means that for $n=1$ the low-energy
Lagrangian of the toy-model reads
\begin{equation}
\mathcal{L}_{\text{le}}(E_{\text{le}},N_{c})=\frac{1}{2}\left(  \partial_{\mu
}\varphi\right)  ^{2}-\frac{1}{2}m^{2}\varphi^{2}+L^{(1)}\varphi^{2}\left(
-\square\right)  \varphi^{2}\text{ ,}%
\end{equation}
where $L^{1}$ is now an unknown parameter. Moreover, the cutoff employed in
the loop function, denoted by $\tilde{\Lambda},$ is also unknown from the
perspective of the low-energy theory. The question is the following: what can
we say from this point of view about the state $S$? Is it possible to fit the
two parameters $L^{(1)}$ and $\tilde{\Lambda}$ in such a way that for
$N_{c}=3$ the approximated curve $T_{BS}(p^{2},1)$ reproduces the `correct' result?

The answer is positive (see Fig. 2, first row). One can reobtain the `bump' of
the $S$ state even in the framework of the low-energy Lagrangian
$\mathcal{L}_{\text{le}},$ in which the field $S$ is not present. What we are
actually doing is a reconstruction of the state $S$: the state $S$ has been
previously integrated out, and then it has been reobtained through a
BS-unitarization study. However, it is clear that the $S$ state, just as the
previously discussed weak $W$ boson, is not a dynamically generated molecular
state of two $\varphi$ fields! This is clear by the way we constructed our
toy-model. However, if one would not know $\mathcal{L}_{\text{toy}}$ but only
the low-energy Lagrangian $\mathcal{L}_{\text{le}}$, one could be led to this
incorrect interpretation of the nature of the $S$ state.
\begin{figure}
[ptb]
\begin{center}
\includegraphics[
height=3.3434in,
width=3.9608in
]%
{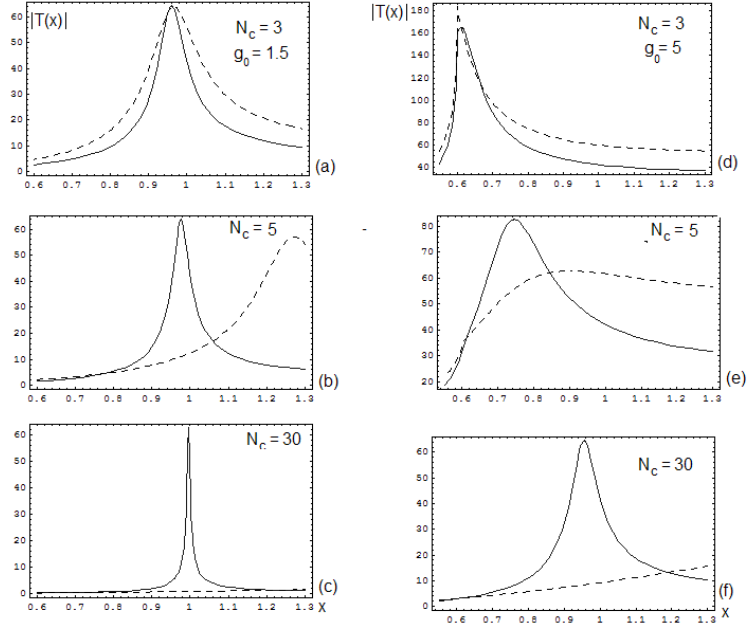}%
\caption{The solid line described the behavior of $\left\vert T(p^{2}%
)\right\vert $ of the $T$-matrix of Eq. (\ref{t-exakt}) for $N_{c}=3$, $5$ and
$30$ and for $g_{0}=1.5$ GeV (left) and $g_{0}=5$ GeV (right) The numerical
values $M_{0}=1$ GeV and $\Lambda=1.5$ GeV have been used. The dashed line
describes the approximated BS matrix $\left\vert T_{BS}(p^{2},1)\right\vert $.
Although for $N_{c}=3$ the two curves are similar, the behavior at large
$N_{c}$ is utterly different: while the solid line becomes (correctly)
narrower for $N_{c}>>1$, the BS-approximation shows the wrong large-$N_{c}$
behavior.}%
\end{center}
\end{figure}

Further comments are in order:

(i) When increasing $N_{c},$ the correct $T$-matrix of Eq. (\ref{t-exakt})
becomes narrower in agreement with the large-$N_{c}$ expectations. On the
contrary, the approximated form $T_{BS}(N_{c},1)$ of Eq. (\ref{tbs}) fades out
in this limit, see Fig. 2. This shows that the BS-inspired unitarization
scheme does not reproduce the correct large-$N_{c}$ result.

(ii) More in general, for each $n$ the correct limit $N_{c}\rightarrow\infty$
cannot be reproduced. This is clear by studying the large-$N_{c}$ limit of Eq.
(\ref{tbs}):
\begin{equation}
T_{BS}(p^{2},n)\overset{N_{c}\rightarrow\infty}{\simeq}-K(n)\text{ .}%
\end{equation}
This result follows from the fact that $K(n)$ scales as $1/N_{c}$ and
$\Sigma_{\Lambda}(p^{2})$ is $N_{c}$-independent. The quantity $K(n)$ is a
polynomial of order $n$ in $p^{2}$ and therefore has no pole for any finite
value of $p^{2}.$ This implies the incorrect result that $M_{S}\rightarrow
\infty$ for $N_{c}\rightarrow\infty.$

(iii) For $n\geq2$ it is possible to recast the BS-scheme in such a way that
the correct large-$N_{c}$ result is obtained \cite{dynrec}. However, this is
not possible for the case $n=1,$ which is generally considered for explicit
hadronic calculations. It is interesting to note that the IAM unitarization
scheme, which is also applicable for $n\geq2,$ is in agreement with the
large-$N_{c}$ results. For a comparative study of different unitarization
schemes see also Ref. \cite{zheng}.

(iv) In the examples of Fig. 2, first row, the required value of $L^{(1)}$ is
for both cases $g_{0}=1.5$ GeV and $g_{0}=5$ GeV close to the correct value
$\frac{g(N_{c})^{2}}{2M_{0}^{4}}.$ However, the required value of the cutoff
$\tilde{\Lambda}$ varies sizably in the two cases: while for $g_{0}=5$ GeV one
has $\tilde{\Lambda}=\Lambda$ (i.e., in agreement with the `correct result')
,in the case $g_{0}=1.5$ GeV one has the unnatural value $\tilde{\Lambda
}=15000\Lambda.$

(v) The cutoff $\tilde{\Lambda}$ of the low-energy version of the toy model
has been taken, just as $\Lambda_{QCD},$ as large-$N_{c}$ independent. This is
indeed a natural choice. However, even including a direct large-$N_{c}$
dependence of $\tilde{\Lambda},$ the qualitative features at large-$N_{c}$ do
not change. This is due to the fact that the loop function depends only
logarithmically on the cutoff.

\subsection{Dynamical reconstruction of the states $f_{2}(1270),$
$a_{1}(1260),$ and $f_{0}(1370)$}

The state $S$ is present as a fundamental, quarkonium-field in the original
toy-model $\mathcal{L}_{\text{toy}}$, it is then integrated out to obtain the
low-energy toy-model $\mathcal{L}_{\text{le}},$ and finally it is reobtained,
i.e. \emph{reconstructed}, via a BS-unitarization of $\mathcal{L}_{\text{le}%
}.$ In this last step it may `looks like' a molecular state of two $\varphi$
fields, however we know that this interpretation is not correct.

The very same interpretation is now proposed for the states $f_{2}(1270),$
$a_{1}(1260),$ and $f_{0}(1370)$: we argue that they are quark-antiquark
fields originally present in the Lagrangian $\mathcal{L}_{had}(E_{\max}%
\simeq2$ GeV$,N_{c}=3);$ the low-energy Lagrangian $\mathcal{L}_{\chi PT+VM}$
is (formally) calculable out of $\mathcal{L}_{had}(E_{\max}\simeq2$
GeV$,N_{c}=3)$ by integrating out the heavier fields, including $f_{2}(1270),$
$a_{1}(1260),$ and $f_{0}(1370).$ This step cannot be performed explicitly
because $\mathcal{L}_{had}(E_{\max}\simeq2GeV,N_{c}=3)$ is not known. Finally,
the states $f_{2}(1270),$ $a_{1}(1260),$ and $f_{0}(1370)$ are reconstructed
out of $\mathcal{L}_{\chi PT+VM}$ using BS-scheme, just as we reconstructed
the $S$ field out of the low-energy Lagrangian $\mathcal{L}_{\text{le}}$ of
the toy model.

From this point of view the predictions of the quark-model and the results of
recent BS-studies agree with each other. It is important to remark that what
is here criticized is not the result of the BS-unitarization, which is a
valuable and correct analysis, but only the related interpretation of the
resonances as hadronic molecular states.

\section{Conclusions}

Some `dynamically generated states' exist for sure: the nuclei. They are
genuine bound state of more fundamental hadrons, the protons and neutrons. The
question discussed in this paper concerns the identification of molecular
states beyond nuclei. To this end we concentrated on the low-energy meson
spectrum and formulated the following question: are the resonances
$f_{2}(1270),$ $a_{1}(1260),$ and $f_{0}(1370)$ dynamically generated
molecular states?

Our answer is negative. With the help of a toy model we have shown that the
interpretation of these states as standard quark-antiquark mesons is
legitimate. In this way there is no conflict between the prediction of the
quark model and the findings of BS unitarizations, which then represent two
alternative ways to describe the same quark-antiquark objects. We believe that
the reconciliation of the quark-model with unitarization studies solves the
following puzzle: the quark-antiquark interpretation works well in the tensor
and axial-vector sectors, to which $f_{2}(1270)$ and $a_{1}(1260)$ belong. If
these resonances would be of different nature, that agreement would have been
-rather surprisingly- accidental.

Future studies are certainly needed. The fact that the quarkonium
interpretation for $f_{2}(1270),$ $a_{1}(1260),$ and $f_{0}(1370)$ is
legitimate, in agreement with present phenomenological information and in a
sense also `desirable', does not represent a conclusive proof.

The description presented in this work is applicable with minor changes also
to other recently investigated mesons between 1-2 GeV (such as the other
members of the tensor and axial-vector nonets) and to dynamically generated
(or reconstructed) states in the baryon and heavy quark sectors.

\end{document}